\documentclass[aip,apl,preprint,a4,preprintnumbers]{revtex4-1}

\usepackage{graphicx}           
\usepackage{dcolumn}            
\usepackage{bm}                 

\begin{document}

\preprint{Preprint}

\title{Efficient and robust fiber coupling of superconducting single photon detectors}

\author{S. N. Dorenbos, R.W. Heeres, E.F.C Driessen and V. Zwiller}
\affiliation{Kavli Institute of Nanoscience, TU Delft, 2628CJ Delft, The Netherlands}

\date{\today}

\begin{abstract}
We applied a recently developed fiber coupling technique to superconducting single photon detectors (SSPDs). As the detector area of SSPDs has to be kept as small as possible, coupling to an optical fiber has been either inefficient or unreliable. Etching through the silicon substrate allows fabrication of a circularly shaped chip which self aligns to the core of a ferrule terminated fiber in a fiber sleeve. In situ alignment at cryogenic temperatures is unnecessary and no thermal stress during cooldown, causing misalignment, is induced. We measured the quantum efficiency of these devices with an attenuated tunable broadband source. The combination of a lithographically defined chip and high precision standard telecommunication components yields near unity coupling efficiency and a system detection efficiency of 34 \% at a wavelength of 1200 nm. This quantum efficiency measurement is confirmed by an absolute efficiency measurement using correlated photon pairs (with $\lambda$ = 1064 nm) produced by spontaneous parametric down-conversion. The efficiency obtained via this method agrees well with the efficiency measured with the attenuated tunable broadband source.
\end{abstract}


\maketitle

At near infrared wavelengths above 1100 nm, i.e. at telecommunication wavelengths, superconducting single photon detectors \cite{Goltsman2001} are an established single photon counting technology. \cite{Hadfield2009} Due to their short dead time\cite{Goltsman2001}, low timing jitter \cite{Tanner2010}, good detection efficiency, and low dark count rate \cite{Dorenbos2008} they have already proved themselves in several experimental applications \cite{Takesue2007,Dorenbos2010,Hadfield2005single}. Furthermore, they are not only suitable for single photon detection, but also for single electron \cite{Rosticher2010} and single plasmon detection \cite{Heeres2010}. For optical experiments efficient coupling to an optical fiber is crucial. Because of the low operating temperature and the small active area of the detector, this task remains challenging. To date different techniques of fiber coupling to SSPDs have been shown. A fiber focuser in combination with a micropositioning stage at low temperature enables in situ alignment. However, because of loss in the fiber focuser and mechanical decoupling of the fiber and the detector, it was estimated that the coupling efficiency was only 80\%. \cite{Hu2009} Another technique which is shown by different groups \cite{Tanner2010,Miki2009} is to mechanically clamp the fiber close to the detector. This can lead to near unity coupling efficiency, but it is very sensitive to small shifts, i.e. due to thermal contraction during cool down. Recently a fiber coupling technique was introduced for transition edge sensor devices.\cite{Miller2011} This technique utilizes micromachining techniques developed for silicon substrates together with optical fiber techniques. Ferrule terminated fibers can be aligned with respect to each other with a commercially available precisely fabricated sleeve. As the dimensions of these sleeves are very precisely defined, it is possible to use sub-micrometer resolution of lithographic technology to fabricate a detector chip with a shape that fits exactly in the sleeve. By fabricating the detector in the middle of the chip, the alignment of the detector with respect to the core of the ferrule terminated fiber follows naturally. 

In figure 1a the device is pictured. The chip, which has a circular shape (figure 1b) is placed inside the fiber sleeve (figure 1a). The detector (figure 1c) is placed exactly in the middle of the circular piece of the chip. The detector itself also has a circular design to minimize the length of the nanowire, as the output of a single mode fiber has a Gaussian shape. Although a smaller active area increases coupling difficulty, it has several advantages. First, a smaller detector yields a shorter recovery time and second a smaller detector has a decreased probability of constrictions, which can severely limit device efficiency.\cite{Kerman2007}  The total length of the nanowire was 847 $\mu$m, which results in a recovery time of approximately 10 ns. We have chosen for an 11 $\mu$m diameter detector to match to the size of a single mode optical fiber (9 $\mu$m). The width of the wire is 100 nm and it has a filling factor of 50\%. The rectangular part of the chip extends outside the fiber ferrule, on which contacts are positioned for bonding. In figure 1d a schematic of the chip is shown. 

\begin{figure}
\centering
\includegraphics{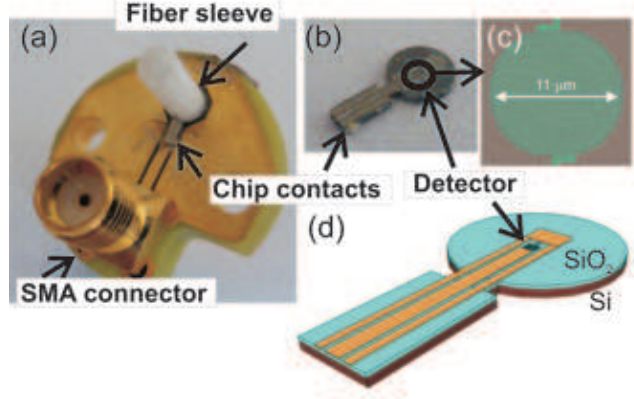}
\caption{\label{fig1} Assembly of the fiber coupled detector. (a) Chip carrier with a mounted chip. To fiber couple the detector, a fiber ferrule is slid into the sleeve. (b) Photograph of chip. (c) Scanning electron microscope picture of the active area of the SSPD. (d) Schematic of the chip.}
\end{figure}

The detector is fabricated by sputtering a thin NbTiN film on oxidized silicon and subsequent electron beam lithography and reactive ion etching.\cite{Dorenbos2008} The circular chip shape is made by deep etching through the silicon substrate. The deep etching is performed by use of the Bosch dry-etch process\cite{Chienlu2005}. This process provides highly anisotropic etching and a vertical etch profile through the 300 $\mu$m thick Si wafer can be obtained. It consists of a longer etching step (with SF$_6$) and a shorter passivation step (with C$_4$F$_8$). The passivation step prevents the side walls from being etched. As a mask for the deep etching a photoresist with a thickness of 5 $\mu$m is placed to protect the SSPD. To be able to apply electron beam lithography for patterning the chip, a triple layer mask is used (photoresist, metal, electron beam resist). After exposure and development of the electron beam resist, the pattern is transferred into the photoresist with reactive ion etching. In this way the flexibility of electron beam lithography is combined with the requirement of a thick deep etch mask. The Bosch process is adjusted such that no measurable undercut or overcut on the bottom side of the chip is observed. The final diameter of the circular part of the chip is measured to be (2.5$\pm$0.002) mm, exactly the size of the ferrule of an FC connector. 

The chip is placed inside a fiber sleeve. The fiber sleeve is glued on a chip carrier, the part of the chip that extends through the opening of the fiber sleeve facilitates the contacts for bonding. After bonding and placing the fiber ferrule inside the connector, the assembly is cooled down by either dipping it into liquid helium or by using a dipstick which allows to cool down to approximately 2.5 K by pumping on the outlet of the dipstick. 

\begin{figure}
\centering
\includegraphics{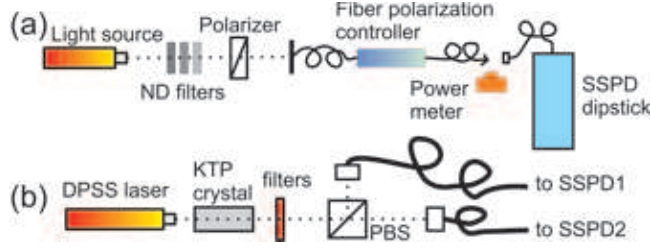}
\caption{\label{fig2} (a) Attenuated tunable broadband setup, consisting of a wavelength filtered white light source, attenuated by neutral density (ND) filters, coupled into a single mode fiber. The polarization is controlled by a polarizer and a fiber coupled polarization controller. Before connecting the fiber to the SSPD, the power in the fiber is measured with a power meter, without the ND filters in place. (b) Setup for absolute quantum efficiency measurement, consisting of a nonlinear crystal, producing collinear type II correlated photon pairs, which are split by a polarizing beam splitter and sent to two detectors.}
\end{figure}

In figure 2a the setup for measuring the efficiency is shown, using an attenuated tunable broadband light source. The light source consists of a white light source with filtering performed by an acoustic-optic modulator. The white light source is a high power ultra-broadband supercontinuum radiation source with a  repetition rate of 20 MHz. The acousto-optic tunable filter system enables selection of a particular wavelength with a linewidth of approximately 6 nm. The setup also consists of different (3) neutral density filters, placed in series in the optical path. The amount of attenuation is measured with a power meter for all the wavelengths of interest. By measuring the neutral density filters separately a precise value of the attenuation is obtained. As the efficiency of the detector is polarization dependent, a fiber coupled polarization controller is used to change the polarization state of the light. The flux of photons in the fiber $N_{ph}$ can be calculated by $N_ph=A \cdot P\lambda/hc$, with P the measured power before attenuation, A the total attenuation and $hc/ \lambda$ the photon energy. The loss at the connection to the power meter is approximated to be equal to the loss at the connection to the detector. The flux of photons in the fiber was between 10$^5$ and 10$^6$ photons/s, much smaller than the repetition rate of the laser and the maximum count rate of the detector, but much larger than the dark count rate of the detector. The system detection efficiency is then defined as the dark count (N$_{dc}$ corrected count rate (N$_c$) divided by the flux of photons in the fiber.
\begin {equation}
SDE =\frac{N_c - N_{dc}}{N_{ph}}
\end{equation}

Another method to measure the quantum efficiency of a detector is based on the detection of correlated photon pairs. This method \cite{Klyshko1977} has previously been used  to characterize avalanche photodiodes \cite{Rarity1987} and solid state photomultipliers \cite{Kwiat1993}. The detection of a photon in a trigger detector heralds the presence of another (correlated) photon, to be detected by the device under investigation. The correlated photon pairs are produced by a nonlinear crystal via the process of spontaneous parametric down-conversion. The setup for this measurement is shown in figure 2b. The crystal is a KTP (KTiOPO$_4$) crystal, from which collinear type-II correlated photon pairs are created. The crystal is pumped with a diode pumped solid state laser, generating photon pairs at a wavelength of $\lambda$=1064 nm with orthogonal polarization. The pair is split at a polarizing beam splitter after filtering out the pump beam. The correlated photons are then sent to two detectors. A time to amplitude converter is used to record the single count rates ($N_{d1}$,$N_{d2}$) and number of correlations ($N_c$). Without accidental and dark counts the efficieny of detector 1 is: $\eta_1 =N_c/N_{d2}$. In the presence of accidental ($R_A$) correlations (caused by dark counts) and a dark count rate ($N_{dc}$) and an imperfect collection efficiency ($p_1$) the equation is the following: 
\begin {equation}
\eta_1 =\frac{N_c-R_A}{p_1 \cdot (N_{d2}-N_{dc})}
\end{equation}
By exchanging the paths, the collection efficiency ($p_1$ and $p_2$) can be determined. 

\begin{figure}
\centering
\includegraphics{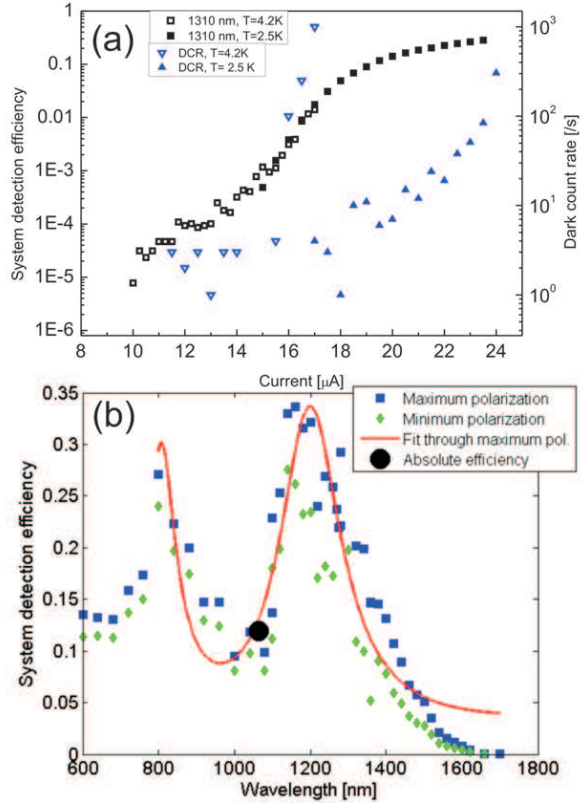}
\caption{\label{fig3}(a) Dark count rate and system detection efficiency (measured with a laser diode with $\lambda=1310$ nm) as a function of bias current for a temperature of 4.2 K and 2.5 K. (b) System detection efficiency as a function of wavelength for polarization, yielding maximum and minimum efficiency. The data for maximum polarization is fitted. The circle indicates the efficiency measured via an absolute efficiency measurement.}
\end{figure}

First, the results using the 'attenuated' method are presented. In figure 3a the dark count rate and system detection efficiency as a function of current through the SSPD are shown for a wavelength of $\lambda$=1310 nm, for the polarization state which gives maximum efficiency. We have shown that the interface between the silicon substrate and the silicon dioxide acts a reflecting surface and maximum absorption can be approximated with $d \approx \lambda/4n + m \cdot \lambda/2n$ with $d$ the thickness and $n$ the refractive index of the silicon dioxide and $m$=0,1,2...\cite{Tanner2010}. With $n=1.5$ and $d=220 nm$, for our device $\lambda_{max}\approx$1300 nm. The system detection efficiency increases exponentially with current towards a plateau, which denotes the highest efficiency for our device. It is clear that at a temperature of 4.2 K this device did not show its maximum efficiency, but decreasing the temperature to $\sim$2.5 K yielded a higher critical current and a higher detection efficiency, with a maximum of 34 \%. The dark count rate as a function of current showed the well known two slope exponential behavior. Background counts caused by detection of stray environment light coupling into the fiber are the major contribution at low currents. For currents close to the critical current the 'intrinsic' dark count rate takes over. The maximum dark count rate was $\sim$ 1000 cps for 4.2 K and $\sim$ 500 cps for 2.5 K. 

In figure 3b the system detection efficiency at 90\% of the critical current (where the dark count rate is approximately 10 cps) is shown as a function of wavelength for the polarization states yielding maximum and minimum efficiency. The operating temperature for this experiment was 2.5 K. The curve shows two peaks in  efficiency with a maximum at a wavelength of 1200 nm, although it has been shown that the intrinsic efficiency of SSPDs is higher for shorter wavelengths. It is apparent that the absorption efficiency plays a large part in the system detection efficiency, as this is highly increased for the cavity specific wavelength of $\lambda$=1200 nm. To explain the shape of the curve we model the optical system according to \cite{Tanner2010} to calculate the absorption efficiency $\eta _{absorption}$. The absorption without optical fiber is simulated in an FDTD simulation, with the optical index ($n,k$) measured with an ellipsometer. To incorporate the effect of the fiber tip the absorption efficiency is multiplied with a function which describes a Fabry-Perot cavity using Fresnel equations. In order to fit the data of figure 3b, we define the system detection efficiency as $SDE=\eta _c \cdot \eta _{intrinsic} \cdot \eta _{absorption}$. For the SDE the coupling efficiency $\eta _c$  is taken as unity. We approximate an exponential wavelength dependence of the intrinsic efficiency $\eta _{intrinsic}$ of the detector. \cite{Tanner2010} 
In figure 3b the resulting fit through the maximum polarization can be seen. The fit explains the shape well. From the fitting we obtain a fiber tip to detector distance of $l$=1.23$ \mu$m, which is a reasonable value as the fiber ferrule is directly in contact with the chip. The fit also yields a maximum value of 33.6\% efficiency at $\lambda$ = 1200 nm.

The results of the 'attenuation' method are verified using the absolute efficiency measurement setup.  We have performed a full set of measurements, with two silicon avalanche photodiodes and another self aligned SSPD as conjugate detector. The coincidences are recorded for a fixed time (120 s). By fitting all data sets, the path and detector efficiencies are determined. The efficiencies of the two Si avalanche photodiodes turn out to be 1.52 \% and 1.57 \%, slightly lower than their specifications (www.optoelectronics.perkinelmer.com). The efficiency of the SSPD under investigation is 11.9 \%. As can be seen in figure 3b (circle) this is in good agreement with the 'attenuation' method.

In conclusion, we have applied a technique which reliably couples a fiber to a superconducting nanowire single photon detector. We have shown a high coupling efficiency and peak system detection efficiency of 34 \% at $\lambda$=1200 nm. In the future we will use fibers with an antireflection coating, which will reduce losses at the end of the fiber. In addition, the optical system can be designed such that it is possible to have the maximum efficiency at any required wavelength by adjusting the thickness of the silicon oxide. \\

\textbf{Acknowledgments}
We acknowledge funding from NWO. We gratefully thank Sae Woo Nam, Adriana Lita and Burm Baek for help and discussions. The etching is performed at the Van Leeuwenhoek Laboratory in Delft with the help of Marc Zuiddam. We thank Gilles Buchs for setting up the tunable broadband source.


\begin{thebibliography}{17}%
\makeatletter
\providecommand \@ifxundefined [1]{%
 \@ifx{#1\undefined}
}%
\providecommand \@ifnum [1]{%
 \ifnum #1\expandafter \@firstoftwo
 \else \expandafter \@secondoftwo
 \fi
}%
\providecommand \@ifx [1]{%
 \ifx #1\expandafter \@firstoftwo
 \else \expandafter \@secondoftwo
 \fi
}%
\providecommand \natexlab [1]{#1}%
\providecommand \enquote  [1]{``#1''}%
\providecommand \bibnamefont  [1]{#1}%
\providecommand \bibfnamefont [1]{#1}%
\providecommand \citenamefont [1]{#1}%
\providecommand \href@noop [0]{\@secondoftwo}%
\providecommand \href [0]{\begingroup \@sanitize@url \@href}%
\providecommand \@href[1]{\@@startlink{#1}\@@href}%
\providecommand \@@href[1]{\endgroup#1\@@endlink}%
\providecommand \@sanitize@url [0]{\catcode `\\12\catcode `\$12\catcode
  `\&12\catcode `\#12\catcode `\^12\catcode `\_12\catcode `\%12\relax}%
\providecommand \@@startlink[1]{}%
\providecommand \@@endlink[0]{}%
\providecommand \url  [0]{\begingroup\@sanitize@url \@url }%
\providecommand \@url [1]{\endgroup\@href {#1}{\urlprefix }}%
\providecommand \urlprefix  [0]{URL }%
\providecommand \Eprint [0]{\href }%
\providecommand \doibase [0]{http://dx.doi.org/}%
\providecommand \selectlanguage [0]{\@gobble}%
\providecommand \bibinfo  [0]{\@secondoftwo}%
\providecommand \bibfield  [0]{\@secondoftwo}%
\providecommand \translation [1]{[#1]}%
\providecommand \BibitemOpen [0]{}%
\providecommand \bibitemStop [0]{}%
\providecommand \bibitemNoStop [0]{.\EOS\space}%
\providecommand \EOS [0]{\spacefactor3000\relax}%
\providecommand \BibitemShut  [1]{\csname bibitem#1\endcsname}%
\let\auto@bib@innerbib\@empty
\bibitem [{\citenamefont {Gol'tsman}\ \emph {et~al.}(2001)\citenamefont
  {Gol'tsman}, \citenamefont {Okunev}, \citenamefont {Chulkova}, \citenamefont
  {Lipatov}, \citenamefont {Semenov}, \citenamefont {Smirnov}, \citenamefont
  {Voronov}, \citenamefont {Dzardanov}, \citenamefont {Williams},\ and\
  \citenamefont {Sobolewski}}]{Goltsman2001}%
  \BibitemOpen
  \bibfield  {author} {\bibinfo {author} {\bibfnamefont {G.}~\bibnamefont
  {Gol'tsman}}, \bibinfo {author} {\bibfnamefont {O.}~\bibnamefont {Okunev}},
  \bibinfo {author} {\bibfnamefont {G.}~\bibnamefont {Chulkova}}, \bibinfo
  {author} {\bibfnamefont {A.}~\bibnamefont {Lipatov}}, \bibinfo {author}
  {\bibfnamefont {A.}~\bibnamefont {Semenov}}, \bibinfo {author} {\bibfnamefont
  {K.}~\bibnamefont {Smirnov}}, \bibinfo {author} {\bibfnamefont
  {B.}~\bibnamefont {Voronov}}, \bibinfo {author} {\bibfnamefont
  {A.}~\bibnamefont {Dzardanov}}, \bibinfo {author} {\bibfnamefont
  {C.}~\bibnamefont {Williams}}, \ and\ \bibinfo {author} {\bibfnamefont
  {R.}~\bibnamefont {Sobolewski}},\ }\href@noop {} {\bibfield  {journal}
  {\bibinfo  {journal} {Applied Physics Letters}\ }\textbf {\bibinfo {volume}
  {79}},\ \bibinfo {pages} {705} (\bibinfo {year} {2001})}\BibitemShut
  {NoStop}%
\bibitem [{\citenamefont {Hadfield}(2009)}]{Hadfield2009}%
  \BibitemOpen
  \bibfield  {author} {\bibinfo {author} {\bibfnamefont {R.~H.}\ \bibnamefont
  {Hadfield}},\ }\href@noop {} {\bibfield  {journal} {\bibinfo  {journal}
  {Nature Photonics}\ }\textbf {\bibinfo {volume} {3}},\ \bibinfo {pages} {696}
  (\bibinfo {year} {2009})}\BibitemShut {NoStop}%
\bibitem [{\citenamefont {Tanner}\ \emph {et~al.}(2010)\citenamefont {Tanner},
  \citenamefont {Natarajan}, \citenamefont {Pottapenjara}, \citenamefont
  {O'Connor}, \citenamefont {Warburton}, \citenamefont {Hadfield},
  \citenamefont {Baek}, \citenamefont {Nam}, \citenamefont {Dorenbos},
  \citenamefont {Bermudez~Urena}, \citenamefont {Zijlstra}, \citenamefont
  {Klapwijk},\ and\ \citenamefont {Zwiller}}]{Tanner2010}%
  \BibitemOpen
  \bibfield  {author} {\bibinfo {author} {\bibfnamefont {M.~G.}\ \bibnamefont
  {Tanner}}, \bibinfo {author} {\bibfnamefont {C.~M.}\ \bibnamefont
  {Natarajan}}, \bibinfo {author} {\bibfnamefont {V.~K.}\ \bibnamefont
  {Pottapenjara}}, \bibinfo {author} {\bibfnamefont {J.~A.}\ \bibnamefont
  {O'Connor}}, \bibinfo {author} {\bibfnamefont {R.~J.}\ \bibnamefont
  {Warburton}}, \bibinfo {author} {\bibfnamefont {R.~H.}\ \bibnamefont
  {Hadfield}}, \bibinfo {author} {\bibfnamefont {B.}~\bibnamefont {Baek}},
  \bibinfo {author} {\bibfnamefont {S.}~\bibnamefont {Nam}}, \bibinfo {author}
  {\bibfnamefont {S.~N.}\ \bibnamefont {Dorenbos}}, \bibinfo {author}
  {\bibfnamefont {E.}~\bibnamefont {Bermudez~Urena}}, \bibinfo {author}
  {\bibfnamefont {T.}~\bibnamefont {Zijlstra}}, \bibinfo {author}
  {\bibfnamefont {T.~M.}\ \bibnamefont {Klapwijk}}, \ and\ \bibinfo {author}
  {\bibfnamefont {V.}~\bibnamefont {Zwiller}},\ }\href@noop {} {\bibfield
  {journal} {\bibinfo  {journal} {Applied Physics Letters}\ }\textbf {\bibinfo
  {volume} {96}},\ \bibinfo {pages} {221109} (\bibinfo {year}
  {2010})}\BibitemShut {NoStop}%
\bibitem [{\citenamefont {Dorenbos}\ \emph {et~al.}(2008)\citenamefont
  {Dorenbos}, \citenamefont {Reiger}, \citenamefont {Perinetti}, \citenamefont
  {Zwiller}, \citenamefont {Zijlstra},\ and\ \citenamefont
  {Klapwijk}}]{Dorenbos2008}%
  \BibitemOpen
  \bibfield  {author} {\bibinfo {author} {\bibfnamefont {S.~N.}\ \bibnamefont
  {Dorenbos}}, \bibinfo {author} {\bibfnamefont {E.~M.}\ \bibnamefont
  {Reiger}}, \bibinfo {author} {\bibfnamefont {U.}~\bibnamefont {Perinetti}},
  \bibinfo {author} {\bibfnamefont {V.}~\bibnamefont {Zwiller}}, \bibinfo
  {author} {\bibfnamefont {T.}~\bibnamefont {Zijlstra}}, \ and\ \bibinfo
  {author} {\bibfnamefont {T.~M.}\ \bibnamefont {Klapwijk}},\ }\href@noop {}
  {\bibfield  {journal} {\bibinfo  {journal} {Applied Physics Letters}\
  }\textbf {\bibinfo {volume} {93}},\ \bibinfo {pages} {131101} (\bibinfo
  {year} {2008})}\BibitemShut {NoStop}%
\bibitem [{\citenamefont {Takesue}\ \emph {et~al.}(2007)\citenamefont
  {Takesue}, \citenamefont {Nam}, \citenamefont {Zhang}, \citenamefont
  {Hadfield}, \citenamefont {Honjo}, \citenamefont {Tamaki},\ and\
  \citenamefont {Yamamoto}}]{Takesue2007}%
  \BibitemOpen
  \bibfield  {author} {\bibinfo {author} {\bibfnamefont {H.}~\bibnamefont
  {Takesue}}, \bibinfo {author} {\bibfnamefont {S.~W.}\ \bibnamefont {Nam}},
  \bibinfo {author} {\bibfnamefont {Q.}~\bibnamefont {Zhang}}, \bibinfo
  {author} {\bibfnamefont {R.~H.}\ \bibnamefont {Hadfield}}, \bibinfo {author}
  {\bibfnamefont {T.}~\bibnamefont {Honjo}}, \bibinfo {author} {\bibfnamefont
  {K.}~\bibnamefont {Tamaki}}, \ and\ \bibinfo {author} {\bibfnamefont
  {Y.}~\bibnamefont {Yamamoto}},\ }\href@noop {} {\bibfield  {journal}
  {\bibinfo  {journal} {Nat Photon}\ }\textbf {\bibinfo {volume} {1}},\
  \bibinfo {pages} {343} (\bibinfo {year} {2007})}\BibitemShut {NoStop}%
\bibitem [{\citenamefont {Dorenbos}\ \emph {et~al.}(2010)\citenamefont
  {Dorenbos}, \citenamefont {Sasakura}, \citenamefont {van Kouwen},
  \citenamefont {Akopian}, \citenamefont {Adachi}, \citenamefont {Namekata},
  \citenamefont {Jo}, \citenamefont {Motohisa}, \citenamefont {Kobayashi},
  \citenamefont {Tomioka}, \citenamefont {Fukui}, \citenamefont {Inoue},
  \citenamefont {Kumano}, \citenamefont {Natarajan}, \citenamefont {Hadfield},
  \citenamefont {Zijlstra}, \citenamefont {Klapwijk}, \citenamefont {Zwiller},\
  and\ \citenamefont {Suemune}}]{Dorenbos2010}%
  \BibitemOpen
  \bibfield  {author} {\bibinfo {author} {\bibfnamefont {S.~N.}\ \bibnamefont
  {Dorenbos}}, \bibinfo {author} {\bibfnamefont {H.}~\bibnamefont {Sasakura}},
  \bibinfo {author} {\bibfnamefont {M.~P.}\ \bibnamefont {van Kouwen}},
  \bibinfo {author} {\bibfnamefont {N.}~\bibnamefont {Akopian}}, \bibinfo
  {author} {\bibfnamefont {S.}~\bibnamefont {Adachi}}, \bibinfo {author}
  {\bibfnamefont {N.}~\bibnamefont {Namekata}}, \bibinfo {author}
  {\bibfnamefont {M.}~\bibnamefont {Jo}}, \bibinfo {author} {\bibfnamefont
  {J.}~\bibnamefont {Motohisa}}, \bibinfo {author} {\bibfnamefont
  {Y.}~\bibnamefont {Kobayashi}}, \bibinfo {author} {\bibfnamefont
  {K.}~\bibnamefont {Tomioka}}, \bibinfo {author} {\bibfnamefont
  {T.}~\bibnamefont {Fukui}}, \bibinfo {author} {\bibfnamefont
  {S.}~\bibnamefont {Inoue}}, \bibinfo {author} {\bibfnamefont
  {H.}~\bibnamefont {Kumano}}, \bibinfo {author} {\bibfnamefont {C.~M.}\
  \bibnamefont {Natarajan}}, \bibinfo {author} {\bibfnamefont {R.~H.}\
  \bibnamefont {Hadfield}}, \bibinfo {author} {\bibfnamefont {T.}~\bibnamefont
  {Zijlstra}}, \bibinfo {author} {\bibfnamefont {T.~M.}\ \bibnamefont
  {Klapwijk}}, \bibinfo {author} {\bibfnamefont {V.}~\bibnamefont {Zwiller}}, \
  and\ \bibinfo {author} {\bibfnamefont {I.}~\bibnamefont {Suemune}},\
  }\href@noop {} {\bibfield  {journal} {\bibinfo  {journal} {Applied Physics
  Letters}\ }\textbf {\bibinfo {volume} {97}},\ \bibinfo {pages} {171106}
  (\bibinfo {year} {2010})}\BibitemShut {NoStop}%
\bibitem [{\citenamefont {Hadfield}\ \emph {et~al.}(2005)\citenamefont
  {Hadfield}, \citenamefont {Stevens}, \citenamefont {Gruber}, \citenamefont
  {Miller}, \citenamefont {Schwall}, \citenamefont {Mirin},\ and\ \citenamefont
  {Nam}}]{Hadfield2005single}%
  \BibitemOpen
  \bibfield  {author} {\bibinfo {author} {\bibfnamefont {R.~H.}\ \bibnamefont
  {Hadfield}}, \bibinfo {author} {\bibfnamefont {M.~J.}\ \bibnamefont
  {Stevens}}, \bibinfo {author} {\bibfnamefont {S.~S.}\ \bibnamefont {Gruber}},
  \bibinfo {author} {\bibfnamefont {A.~J.}\ \bibnamefont {Miller}}, \bibinfo
  {author} {\bibfnamefont {R.~E.}\ \bibnamefont {Schwall}}, \bibinfo {author}
  {\bibfnamefont {R.~P.}\ \bibnamefont {Mirin}}, \ and\ \bibinfo {author}
  {\bibfnamefont {S.~W.}\ \bibnamefont {Nam}},\ }\href@noop {} {\bibfield
  {journal} {\bibinfo  {journal} {Opt. Express}\ }\textbf {\bibinfo {volume}
  {13}},\ \bibinfo {pages} {10846} (\bibinfo {year} {2005})}\BibitemShut
  {NoStop}%
\bibitem [{\citenamefont {Rosticher}\ \emph {et~al.}(2010)\citenamefont
  {Rosticher}, \citenamefont {Ladan}, \citenamefont {Maneval}, \citenamefont
  {Dorenbos}, \citenamefont {Zijlstra}, \citenamefont {Klapwijk}, \citenamefont
  {Zwiller}, \citenamefont {Lupacu},\ and\ \citenamefont
  {Nogues}}]{Rosticher2010}%
  \BibitemOpen
  \bibfield  {author} {\bibinfo {author} {\bibfnamefont {M.}~\bibnamefont
  {Rosticher}}, \bibinfo {author} {\bibfnamefont {F.~R.}\ \bibnamefont
  {Ladan}}, \bibinfo {author} {\bibfnamefont {J.~P.}\ \bibnamefont {Maneval}},
  \bibinfo {author} {\bibfnamefont {S.~N.}\ \bibnamefont {Dorenbos}}, \bibinfo
  {author} {\bibfnamefont {T.}~\bibnamefont {Zijlstra}}, \bibinfo {author}
  {\bibfnamefont {T.~M.}\ \bibnamefont {Klapwijk}}, \bibinfo {author}
  {\bibfnamefont {V.}~\bibnamefont {Zwiller}}, \bibinfo {author} {\bibfnamefont
  {A.}~\bibnamefont {Lupacu}}, \ and\ \bibinfo {author} {\bibfnamefont
  {G.}~\bibnamefont {Nogues}},\ }\href@noop {} {\bibfield  {journal} {\bibinfo
  {journal} {Applied Physics Letters}\ }\textbf {\bibinfo {volume} {97}}
  (\bibinfo {year} {2010})}\BibitemShut {NoStop}%
\bibitem [{\citenamefont {Heeres}\ \emph {et~al.}(2010)\citenamefont {Heeres},
  \citenamefont {Dorenbos}, \citenamefont {Koene}, \citenamefont {Solomon},
  \citenamefont {Kouwenhoven},\ and\ \citenamefont {Zwiller}}]{Heeres2010}%
  \BibitemOpen
  \bibfield  {author} {\bibinfo {author} {\bibfnamefont {R.~W.}\ \bibnamefont
  {Heeres}}, \bibinfo {author} {\bibfnamefont {S.~N.}\ \bibnamefont
  {Dorenbos}}, \bibinfo {author} {\bibfnamefont {B.}~\bibnamefont {Koene}},
  \bibinfo {author} {\bibfnamefont {G.~S.}\ \bibnamefont {Solomon}}, \bibinfo
  {author} {\bibfnamefont {L.~P.}\ \bibnamefont {Kouwenhoven}}, \ and\ \bibinfo
  {author} {\bibfnamefont {V.}~\bibnamefont {Zwiller}},\ }\href@noop {}
  {\bibfield  {journal} {\bibinfo  {journal} {Nano Letters}\ }\textbf {\bibinfo
  {volume} {10}},\ \bibinfo {pages} {661} (\bibinfo {year} {2010})}\BibitemShut
  {NoStop}%
\bibitem [{\citenamefont {Hu}\ \emph {et~al.}(2009)\citenamefont {Hu},
  \citenamefont {Zhong}, \citenamefont {White}, \citenamefont {Dauler},
  \citenamefont {Najafi}, \citenamefont {Herder}, \citenamefont {Wong},\ and\
  \citenamefont {Berggren}}]{Hu2009}%
  \BibitemOpen
  \bibfield  {author} {\bibinfo {author} {\bibfnamefont {X.}~\bibnamefont
  {Hu}}, \bibinfo {author} {\bibfnamefont {T.}~\bibnamefont {Zhong}}, \bibinfo
  {author} {\bibfnamefont {J.~E.}\ \bibnamefont {White}}, \bibinfo {author}
  {\bibfnamefont {E.~A.}\ \bibnamefont {Dauler}}, \bibinfo {author}
  {\bibfnamefont {F.}~\bibnamefont {Najafi}}, \bibinfo {author} {\bibfnamefont
  {C.~H.}\ \bibnamefont {Herder}}, \bibinfo {author} {\bibfnamefont {F.~N.~C.}\
  \bibnamefont {Wong}}, \ and\ \bibinfo {author} {\bibfnamefont {K.~K.}\
  \bibnamefont {Berggren}},\ }\href@noop {} {\bibfield  {journal} {\bibinfo
  {journal} {Opt. Lett.}\ }\textbf {\bibinfo {volume} {34}},\ \bibinfo {pages}
  {3607} (\bibinfo {year} {2009})}\BibitemShut {NoStop}%
\bibitem [{\citenamefont {Miki}\ \emph {et~al.}(2009)\citenamefont {Miki},
  \citenamefont {Takeda}, \citenamefont {Fujiwara}, \citenamefont {Sasaki},\
  and\ \citenamefont {Wang}}]{Miki2009}%
  \BibitemOpen
  \bibfield  {author} {\bibinfo {author} {\bibfnamefont {S.}~\bibnamefont
  {Miki}}, \bibinfo {author} {\bibfnamefont {M.}~\bibnamefont {Takeda}},
  \bibinfo {author} {\bibfnamefont {M.}~\bibnamefont {Fujiwara}}, \bibinfo
  {author} {\bibfnamefont {M.}~\bibnamefont {Sasaki}}, \ and\ \bibinfo {author}
  {\bibfnamefont {Z.}~\bibnamefont {Wang}},\ }\href@noop {} {\bibfield
  {journal} {\bibinfo  {journal} {Opt. Express}\ }\textbf {\bibinfo {volume}
  {17}},\ \bibinfo {pages} {23557} (\bibinfo {year} {2009})}\BibitemShut
  {NoStop}%
\bibitem [{\citenamefont {Miller}\ \emph {et~al.}(2011)\citenamefont {Miller},
  \citenamefont {Lita}, \citenamefont {Calkins}, \citenamefont {Vayshenker},
  \citenamefont {Gruber},\ and\ \citenamefont {Nam}}]{Miller2011}%
  \BibitemOpen
  \bibfield  {author} {\bibinfo {author} {\bibfnamefont {A.~J.}\ \bibnamefont
  {Miller}}, \bibinfo {author} {\bibfnamefont {A.~E.}\ \bibnamefont {Lita}},
  \bibinfo {author} {\bibfnamefont {B.}~\bibnamefont {Calkins}}, \bibinfo
  {author} {\bibfnamefont {I.}~\bibnamefont {Vayshenker}}, \bibinfo {author}
  {\bibfnamefont {S.~M.}\ \bibnamefont {Gruber}}, \ and\ \bibinfo {author}
  {\bibfnamefont {S.~W.}\ \bibnamefont {Nam}},\ }\href@noop {} {\bibfield
  {journal} {\bibinfo  {journal} {Opt. Express}\ }\textbf {\bibinfo {volume}
  {19}},\ \bibinfo {pages} {9102} (\bibinfo {year} {2011})}\BibitemShut
  {NoStop}%
\bibitem [{\citenamefont {Kerman}\ \emph {et~al.}(2007)\citenamefont {Kerman},
  \citenamefont {Dauler}, \citenamefont {Yang}, \citenamefont {Rosfjord},
  \citenamefont {Anant}, \citenamefont {Berggren}, \citenamefont {Gol'tsman},\
  and\ \citenamefont {Voronov}}]{Kerman2007}%
  \BibitemOpen
  \bibfield  {author} {\bibinfo {author} {\bibfnamefont {A.~J.}\ \bibnamefont
  {Kerman}}, \bibinfo {author} {\bibfnamefont {E.~A.}\ \bibnamefont {Dauler}},
  \bibinfo {author} {\bibfnamefont {J.~K.~W.}\ \bibnamefont {Yang}}, \bibinfo
  {author} {\bibfnamefont {K.~M.}\ \bibnamefont {Rosfjord}}, \bibinfo {author}
  {\bibfnamefont {V.}~\bibnamefont {Anant}}, \bibinfo {author} {\bibfnamefont
  {K.~K.}\ \bibnamefont {Berggren}}, \bibinfo {author} {\bibfnamefont {G.~N.}\
  \bibnamefont {Gol'tsman}}, \ and\ \bibinfo {author} {\bibfnamefont {B.~M.}\
  \bibnamefont {Voronov}},\ }\href@noop {} {\bibfield  {journal} {\bibinfo
  {journal} {Applied Physics Letters}\ }\textbf {\bibinfo {volume} {90}},\
  \bibinfo {pages} {101110} (\bibinfo {year} {2007})}\BibitemShut {NoStop}%
\bibitem [{\citenamefont {Chienliu}(2005)}]{Chienlu2005}%
  \BibitemOpen
  \bibfield  {author} {\bibinfo {author} {\bibfnamefont {C.}~\bibnamefont
  {Chienliu}},\ }\href@noop {} {\bibfield  {journal} {\bibinfo  {journal}
  {Journal of Micromechanics and Microengineering}\ }\textbf {\bibinfo {volume}
  {15}},\ \bibinfo {pages} {580} (\bibinfo {year} {2005})}\BibitemShut
  {NoStop}%
\bibitem [{\citenamefont {Klyshko}(1977)}]{Klyshko1977}%
  \BibitemOpen
  \bibfield  {author} {\bibinfo {author} {\bibfnamefont {D.}~\bibnamefont
  {Klyshko}},\ }\href@noop {} {\bibfield  {journal} {\bibinfo  {journal} {Sov.
  J. Quantum Electron.}\ }\textbf {\bibinfo {volume} {7}},\ \bibinfo {pages}
  {591} (\bibinfo {year} {1977})}\BibitemShut {NoStop}%
\bibitem [{\citenamefont {Rarity}, \citenamefont {Ridley},\ and\ \citenamefont
  {Tapster}(1987)}]{Rarity1987}%
  \BibitemOpen
  \bibfield  {author} {\bibinfo {author} {\bibfnamefont {J.~G.}\ \bibnamefont
  {Rarity}}, \bibinfo {author} {\bibfnamefont {K.~D.}\ \bibnamefont {Ridley}},
  \ and\ \bibinfo {author} {\bibfnamefont {P.~R.}\ \bibnamefont {Tapster}},\
  }\href@noop {} {\bibfield  {journal} {\bibinfo  {journal} {Appl. Opt.}\
  }\textbf {\bibinfo {volume} {26}},\ \bibinfo {pages} {4616} (\bibinfo {year}
  {1987})}\BibitemShut {NoStop}%
\bibitem [{\citenamefont {Kwiat}\ \emph {et~al.}(1993)\citenamefont {Kwiat},
  \citenamefont {Steinberg}, \citenamefont {Chiao}, \citenamefont {Eberhard},\
  and\ \citenamefont {Petroff}}]{Kwiat1993}%
  \BibitemOpen
  \bibfield  {author} {\bibinfo {author} {\bibfnamefont {P.~G.}\ \bibnamefont
  {Kwiat}}, \bibinfo {author} {\bibfnamefont {A.~M.}\ \bibnamefont
  {Steinberg}}, \bibinfo {author} {\bibfnamefont {R.~Y.}\ \bibnamefont
  {Chiao}}, \bibinfo {author} {\bibfnamefont {P.~H.}\ \bibnamefont {Eberhard}},
  \ and\ \bibinfo {author} {\bibfnamefont {M.~D.}\ \bibnamefont {Petroff}},\
  }\href@noop {} {\bibfield  {journal} {\bibinfo  {journal} {Physical Review
  A}\ }\textbf {\bibinfo {volume} {48}},\ \bibinfo {pages} {R867} (\bibinfo
  {year} {1993})}\BibitemShut {NoStop}%
\end{thebibliography}
%

\end{document}